\documentclass[aps,showpacs, superscriptaddress,preprint,preprintnumbers,amsmath,amssymb]{revtex4}
\usepackage{graphicx}
\usepackage{dcolumn}
\usepackage{bm}
\begin{document}
\renewcommand{\thefootnote}{\fnsymbol{footnote}}
\title{Non-saturating Quantum Magnetization in Weyl semimetal TaAs}

\author{Cheng-Long Zhang$^\star$}
\affiliation{International Center for Quantum Materials, School of Physics, Peking University, Beijing 100871, China}

\author{C. M. Wang$^\star$}
\affiliation{Department of Physics, South University of Science and Technology of China, Shenzhen 518055, China}
\affiliation{Shenzhen Key Laboratory of Quantum Science and Engineering, Shenzhen 518055, China}
\affiliation{School of Physics and Electrical Engineering, Anyang Normal University, Anyang 455000, China}

\author{Zhujun Yuan}
\affiliation{International Center for Quantum Materials, School of Physics, Peking University, Beijing 100871, China}

\author{Chi-Cheng Lee}
\affiliation{Centre for Advanced 2D Materials and Graphene Research Centre National University of Singapore, 6 Science Drive 2, Singapore 117546}
\affiliation{Department of Physics, National University of Singapore, 2 Science Drive 3, Singapore 117542}

\author{Li Pi}
\affiliation{High Magnetic Field Laboratory, Chinese Academy of Sciences, Hefei 230031, Anhui, China}
\affiliation{Hefei Science Center, Chinese Academy of Sciences, Hefei 230031, Anhui, China}

\author{Changying Xi}
\affiliation{High Magnetic Field Laboratory, Chinese Academy of Sciences, Hefei 230031, Anhui, China}
\affiliation{Hefei Science Center, Chinese Academy of Sciences, Hefei 230031, Anhui, China}

\author{Hsin Lin}
\affiliation{Centre for Advanced 2D Materials and Graphene Research Centre National University of Singapore, 6 Science Drive 2, Singapore 117546}
\affiliation{Department of Physics, National University of Singapore, 2 Science Drive 3, Singapore 117542}

\author{Neil Harrison}
\affiliation{National High Magnetic Field Laboratory, Los Alamos National Laboratory, MS E536, Los Alamos, NM 87545, USA}

\author{Hai-Zhou Lu}
\email{luhz@sustc.edu.cn}
\affiliation{Department of Physics, South University of Science and Technology of China, Shenzhen 518055, China}
\affiliation{Shenzhen Key Laboratory of Quantum Science and Engineering, Shenzhen 518055, China}

\author{Jinglei Zhang}
\email{zhangjinglei@hmfl.ac.cn}
\affiliation{High Magnetic Field Laboratory, Chinese Academy of Sciences, Hefei 230031, Anhui, China}
\affiliation{Hefei Science Center, Chinese Academy of Sciences, Hefei 230031, Anhui, China}

\author{Shuang Jia}
\email{gwljiashuang@pku.edu.cn}
\affiliation{International Center for Quantum Materials, School of Physics, Peking University, Beijing 100871, China}
\affiliation{Collaborative Innovation Center of Quantum Matter, Beijing 100871, China}

\clearpage

\begin{abstract}
Detecting the spectroscopic signatures of Dirac-like quasiparticles in emergent topological materials is crucial for searching their potential applications.
Magnetometry is a powerful tool for fathoming electrons in solids, yet its ability for discerning Dirac-like quasiparticles has not been recognized.
Adopting the probes of magnetic torque and parallel magnetization for the archetype Weyl semimetal TaAs in strong magnetic field,
we observed a quasi-linear field dependent effective transverse magnetization and a strongly enhanced parallel magnetization when the system is in the quantum limit. Distinct from the saturating magnetic responses for massive carriers, the non-saturating signals of TaAs in strong field is consistent with our newly developed magnetization calculation for a Weyl fermion system in an arbitrary angle.  Our results for the first time establish a thermodynamic criterion for detecting the unique magnetic response of 3D massless Weyl fermions in the quantum limit.

\end{abstract}
\maketitle

The low-energy states of electrons in topological materials can be described as a series of quasiparticles which obey different representations of the Dirac equation \cite{Basov14rmp, Hasan10rmp, Charlier07rmp,Novoselov12nat, Liang15nmat}. One kind of the three-dimensional (3D) massless quasiparticle is Weyl fermion, which has been discovered in topological Weyl and Dirac semimetals \cite{Wang12prb, Wang13prb, Liu14sci, Liu14natmat, Huang15nc, Weng15prx, Xu15sci-TaAs, Lv15prx}. The Weyl quasiparticles occur in the vicinity of a finite number of band touching points, dubbed Weyl nodes, in these topological semimetals. The unique topological nature of the Weyl semimetal promises many novel properties belonging to the massless quasiparticles, such as linear energy dispersion, monopoles and Fermi arcs on the surface \cite{Wan11prb, Yang11prb, Burkov11prl}.Of particular, the 0th Landau bands (LBs) of the Weyl fermions in strong magnetic field are purely chiral modes \cite{Adler69pr, Bell69Jackiw,Nielsen83plb}. These one-dimensional (1D), chiral LBs are expected to exhibit a negative longitudinal magnetoresistance (MR) as a signature of the long-sought chiral anomaly in quantum field theory \cite{Son13prb}. Realizing the chiral anomaly in solids has inspired intensive experimental activities on topological semimetals in strong magnetic field \cite{Kim13prl, Li16np, LiCZ15nc, LiH16nc, ZhangC17nc, HuangXC15prx, ZhangCL16nc}. Nevertheless, the MR in the quantum limit (QL) also depends sophisticatedly on the nature of the impurity scattering \cite{WangCM16prl} and thus it cannot give the information of the quasiparticles' spectrum deterministically \cite{Lu15prb-QL,Goswami15prb,ZhangSB16njp}. Actually the negative MR has not been observed experimentally in the QL of topological semimetals.

By contrast, the magnetic responses of the electrons are much less complex, because they do not interplay with impurity scattering. Indeed they are simply determined by the derivatives of the electrons' thermodynamical potential $\Omega$ with respect to magnetic field $H$, and they have been used to probe the properties of the Fermi surface, including the topological aspect \cite{Shoenberg84book, Sebastian08nat, Li08sci}.
Previous studies claimed that the non-trivial Berry phase account for the non-zero extrapolation of the quantum oscillations in topological semimetals \cite{Xiao10rmp, Goodrich02prl, Mikitik04prl, Moll16nc, WangCM16prl}. However a generic band structure in a gapped semimetal such as ZrTe$_5$ can also carry non-zero Berry curvature where the massive carriers occupy the bottom of the conduction band \cite{zrte5}. In this paper we suggest a deterministic criterion of the magnetic response for massless topological electrons. We focus on the magnetic response of the 0th LB in a sufficiently strong magnetic field in which all of the rest LBs have left the $E_F$. In such extreme gapped and crossing bands show sheerly different magnetic response. To illustrate the difference, we choose TaAs as a prototype topological semimetal hosting well-defined Weyl quasi-electrons. The cross section area of the Weyl pocket is sufficiently small so that a steady magnetic field can approach the QL within a large deviation of the angle in a magnetic torque measurement. We show that the effective transverse ($M_T$) and parallel magnetizations ($M_{||}$) of TaAs are quasi-linear field dependent beyond the QL. Consistent with our calculations, these non-saturating magnetic responses are distinct from that for massive electrons, which can serve as a thermodynamic criterion for discerning the massless particles in emergent topological materials.

\begin{figure}
  \includegraphics[clip, width=1\textwidth]{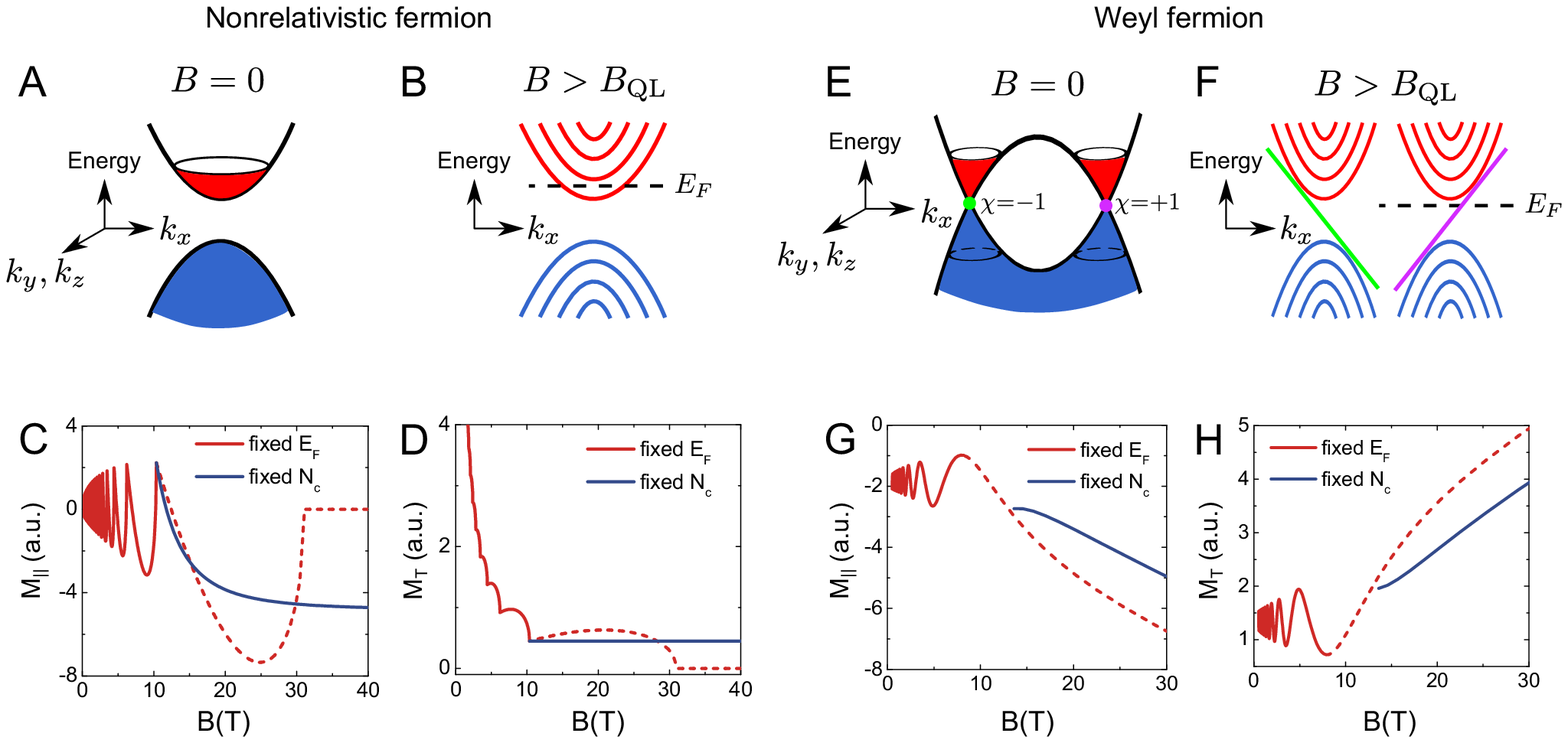}\\
  \caption{ \textbf{Magnetic responses of the nonrelativistic and Weyl fermions.} (A) The energy bands of nonrelativistic fermions in zero magnetic field. (B) The LBs of nonrelativistic fermions form in a magnetic field. Only the 0th LB crosses the $E_F$ in the QL. (C) and (D): Calculated parallel  magnetization ($M_{||}$) and effective transverse magnetization ($M_T$) of nonrelativistic fermions as functions of magnetic field. We used two constraints for the calculation in the QL: fixed the $E_F$ (red line) and fixed carrier density ($N_c$) (blue line). (E) A typical schematic energy bands of a pair of type I Weyl nodes in zero magnetic field. (F) A series of LBs for a pair of Weyl nodes form in a magnetic field. Two 0th LBs are entirely chiral (green and violet). (G) and (H): Calculated $M_{||}$ and $M_T$ as  functions of magnetic field, respectively.}
  \label{Fig1}
\end{figure}

Before discussing the experimental results, we firstly depict the pictures of the magnetic response for a Weyl fermion system and a nonrelativistic electron system comparably (see following text and SI for the details). Figure~\ref{Fig1}\textbf{A} and \textbf{B} are the sketches of the bands for the nonrelativistic electrons and holes in a magnetic field, in which the magnetization is contributed by either electrons or holes which are separated by a band gap. When the field increases, the LBs successively leave the $E_F$ (Fig.~\ref{Fig1}\textbf{B}),  leading to the oscillatory $M_{||}$ around zero and the dropping $M_T$ with respect to field (Fig.~\ref{Fig1}\textbf{C} and \textbf{D}) until all of the LBs have left the $E_F$ except for the lowest one.
The $E_F$ remains intact in low field when there are a lot of LBs below it. However the change of the $E_F$  in strong magnetic field cannot be ignored when the system enters the QL (see SI). Here we consider two alternative constraints: imposing the conservation of the carrier concentration ($N_c$) or fixing the $E_F$.
If the $E_F$ is fixed, the zero-point energy of the 0th LB will be lifted by a critical field higher than the $E_F$, which will vanish both $M_{||}$ and $M_T$ (red dot lines in Figs.~\ref{Fig1}\textbf{C} and \textbf{D}). If the $N_c$ is fixed,  the $E_F$ will be pinned at the edge of the 0th LB regardless the magnetic field increase. Under this constrain, $M_{||}$ will saturate and $M_T$ will be a constant (blue lines in Fig.~\ref{Fig1}\textbf{C} and \textbf{D}). In other words, both $M_{||}$ and $M_T$ for nonrelativistic electrons are invariant in the QL due to the existence of the band gap.  The profile of the saturated $M_{||}$ and $M_T$ in the QL has been observed in 3D massive (gapped) bulk systems such as Bi, sulfur-doped Bi$_2$Te$_3$ and InSb \cite{Wangyayu,Li08sci,Brignall74jpc}. By contrast, gapless Weyl fermions contribute both positive and negative energy bands to the magnetization (Fig.~\ref{Fig1}\textbf{E}), leading to an oscillatory $M_{||}$ and $M_T$ around a mean value in low magnetic fields (Figs.~\ref{Fig1}\textbf{G} and \textbf{H}).
When the rest of the LBs have left the $E_F$, both chiral modes will contribute to the magnetic response (Fig.~\ref{Fig1}\textbf{F}), leading to non-saturating $M_{||}$ and $M_T$ in strong magnetic field. This essential difference of the magnetic responses in the QL stems from their topological nature.

\begin{figure}[h!]
  \includegraphics[clip, width=0.9\textwidth]{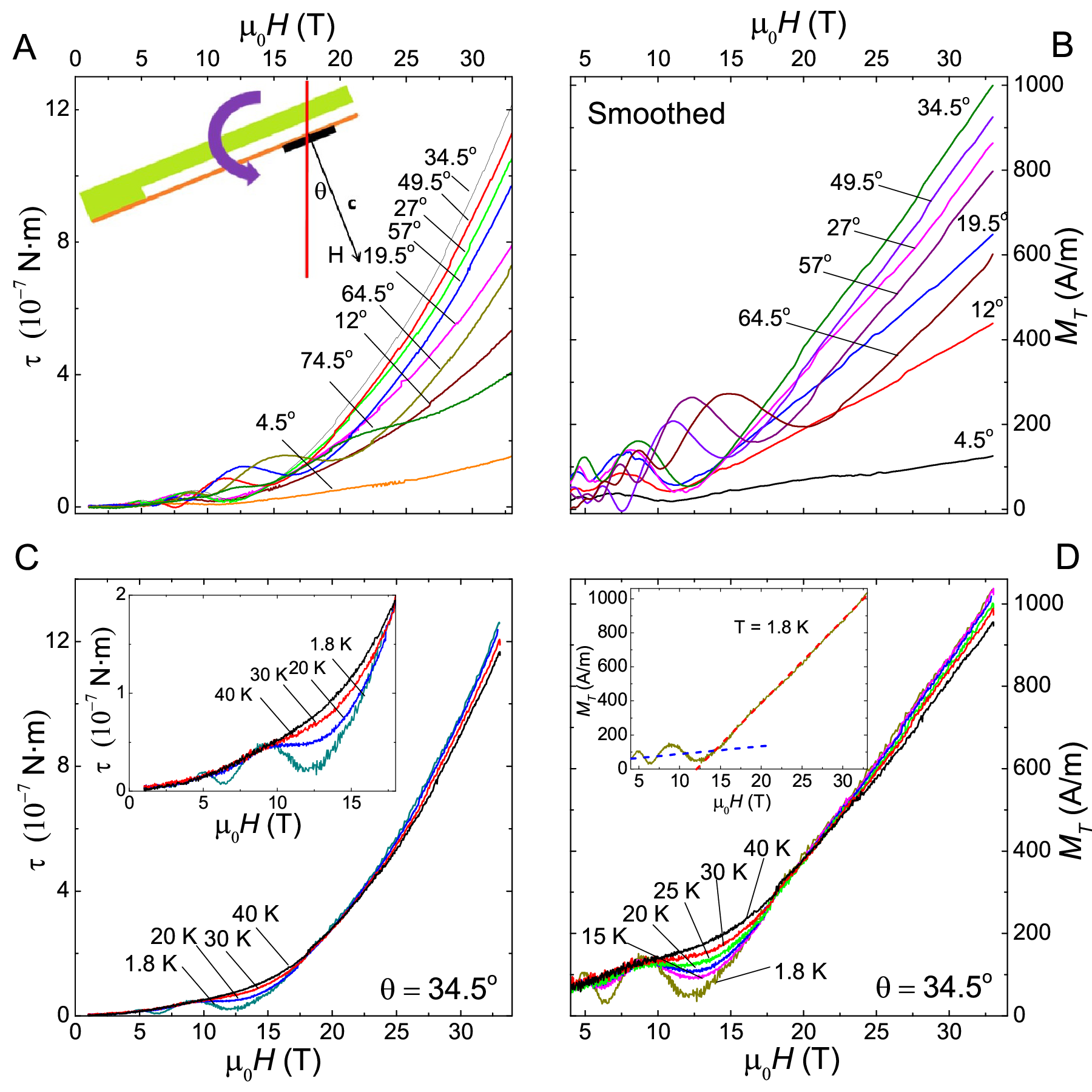}\\
  \caption{ \textbf{Measured $\tau$ and $M_T$ of TaAs versus magnetic field at different temperatures and angles.} (A) and (B) $\tau$ and $M_T$ at 1.8 K at different tilted angles, respectively. Inset in (A) shows the rotation setup where the angle $\theta$ is defined as the tilt off the c axis. The curves in (B) have been smoothed. (C) and (D) $\tau$ and $M_T$ in a fixed angle ($\theta$ = 34.5$^\circ$) at different temperatures, respectively. Inset in (C) shows a zoom-in in low fields where strong temperature dependent dHvA oscillations superpose on a parabolic background. Inset in (D) shows the $M_T$ curve at 1.8 K, where two dashed colored lines show the low- and high-field slopes. The slope takes a obvious enhancement near the QL.}
  \label{Fig2}
\end{figure}

The depiction above sheds light for understanding the magnetic torque signals of TaAs. The magnetic torque $\tau$ is defined as $-\frac{\partial{\Omega}}{\partial{\theta}}$, where $\theta$ is the angle of the magnetic field with respect to the \textbf{c}-axis. In a macroscopic expression $\tau$ is formulated as a mechanic torsional torque, $\tau$ = V$\textbf{M}\times\mu_0\textbf{H}$ = $\mu_0$VHM$_T$. Figure~\ref{Fig2}\textbf{A} and \textbf{B} show $\tau$ and $M_T$ against magnetic field at several representative tilted angles, respectively. Clear de Haas--van Alphen (dHvA) oscillations superpose on a large diamagnetic background at different temperatures. In essence, only one frequency is detected at small angles and no dHvA oscillations are found beyond the QL of these quantum oscillations. Instead, $M_T$ linearly increases with respect to H, with a pronounced slope change after entering the QL.
The change of the dHvA oscillations and the magnetic response at different temperatures for a tilt angle $\theta$ = 34.5$^\circ$ are shown in Fig.~\ref{Fig2}\textbf{C} and \textbf{D}, respectively. The dHvA oscillations decay at higher temperatures while $\tau$ and $M_T$  remain intact beyond the QL. It is noteworthy that the extrapolated intercept of the linear $M_T$ is far less than zero, which indicates that this featureless $M_T$ in strong magnetic fields is not the same as the commonly observed magnetization proportional to low fields. We mention that $M_T$ for TaAs in the QL is quite different from that for NbAs \cite{Moll16nc}, where the high-field torque signal is linear with respect to field in the QL and then bends over at 50 T. The linear $\tau$ corresponds to a saturating $M_T$, which is similar to that in sulfur-doped Bi$_2$Te$_3$ \cite{Wangyayu}.



\begin{figure}[h!]
  \includegraphics[clip, width=0.4\textwidth]{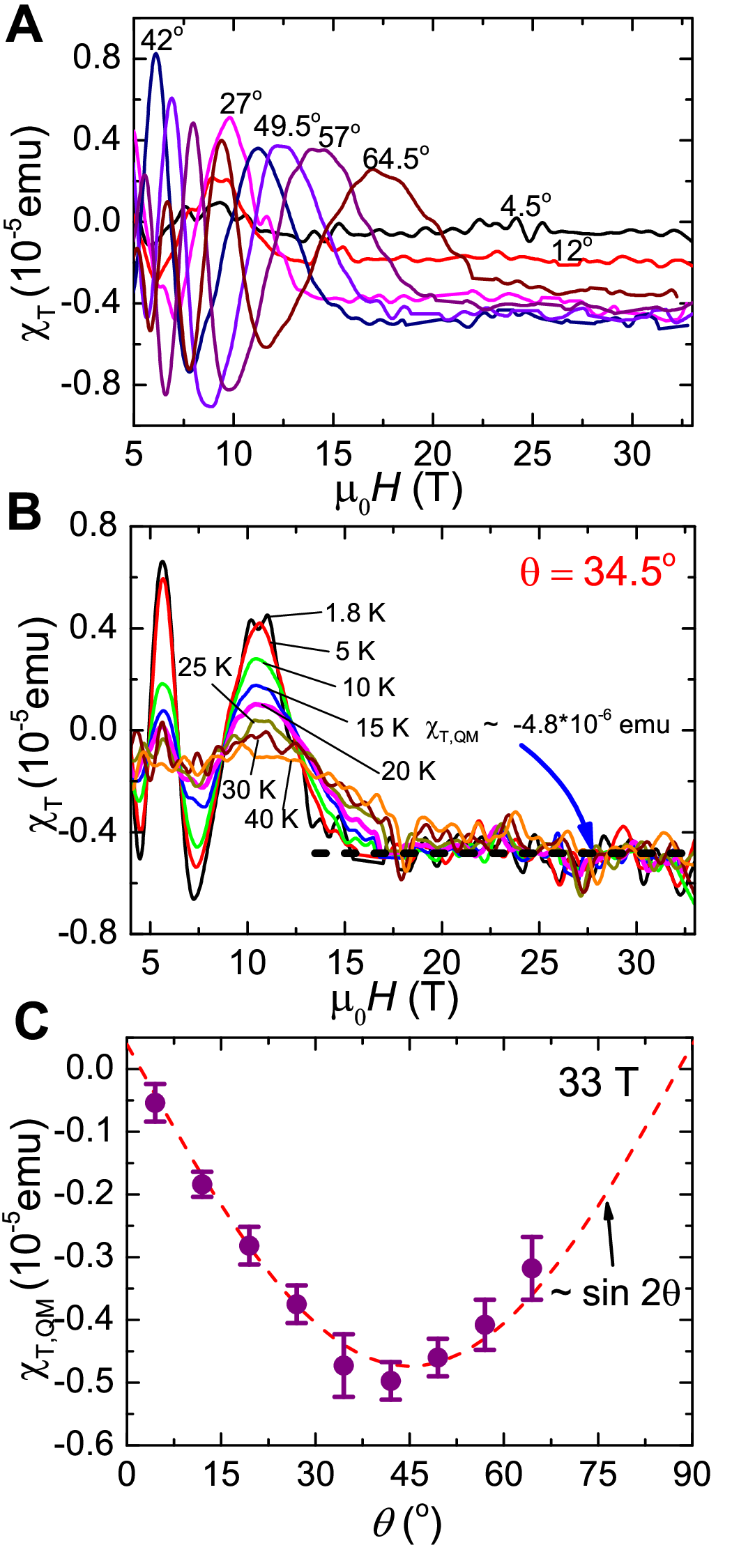}\\
 \caption{\textbf{Differential effective susceptibility ($\chi_T$) for TaAs.} (A) $\chi_T$ at different angles. (B) $\chi_T$ for $\theta$ = 34.5$^\circ$ at different temperatures. (C) The heights of the $\chi_T$ plateaus at 33 T versus angles. Red dashed line shows the fitting with a relation of $\sin{2\theta}$.    }
  \label{Fig3}
\end{figure}

\begin{figure}[h!]
  \includegraphics[clip, width=0.5\textwidth]{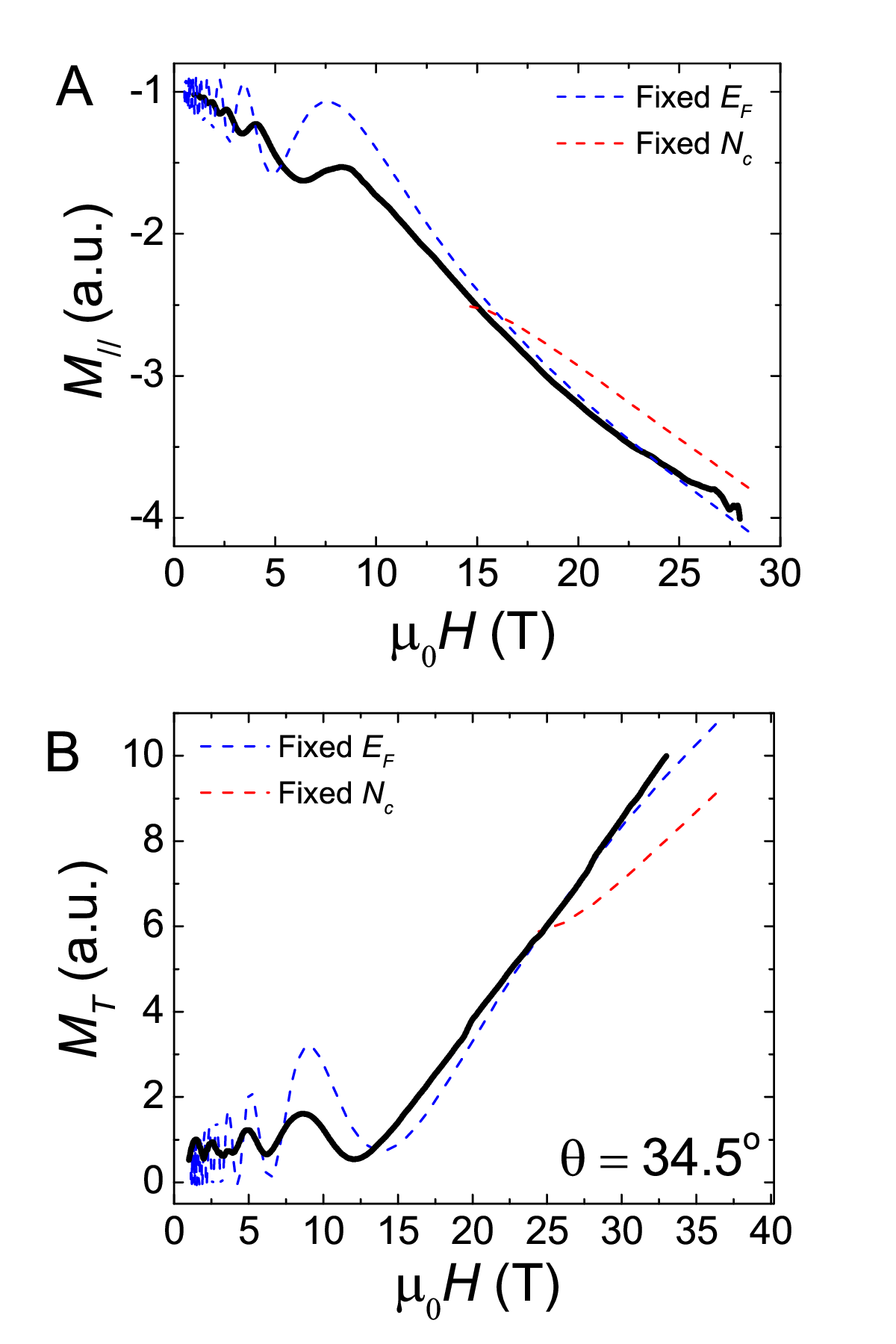}\\
  \caption{\textbf{Comparison between experiments and theory for the $M_\parallel$ and $M_T$ of TaAs. } (A) $M_\parallel$ (black line) and (B) $M_T$ (black line). The blue dashed lines represent the theoretical results with the constraint of the fixed the $E_F$ in the full range of magnetic field, while the red dashed lines represent the results with the constraint of the fixed $N_c$ in strong fields. }
  \label{Fig4}
\end{figure}


To understand the non-saturating $M_T$, we plot the differential susceptibility $\chi_T$ = $\partial$$M_T$/$\partial$$H$ at different angles and temperatures with respect to the field in Figs.~\ref{Fig3}\textbf{A} and \textbf{B}, respectively. With increasing field, the $\chi_T$ evolves from quantum oscillations to a plateau when the QL is achieved at different angles. The height of the plateaus remains intact with increasing temperatures, which is sheerly different from the damping amplitude of the dHvA oscillations in lower fields (Fig.~\ref{Fig3}\textbf{B}). Fig.~\ref{Fig3}\textbf{C} shows that the heights of the plateaus above the QL at different angles can be well-fitted by a relation of $\sin{2\theta}$ for $\theta$ $<$ 60$^\circ$, consistent with the expression of $\tau$ = V$\textbf{M}\times\mu_0\textbf{H}$ which can be reformulated as $\frac{1}{2}\Delta\chi\mu_0H^2V\sin{2\theta}$ ($\Delta\chi$ = $\chi_\parallel-\chi_\perp$).

Now we show that the magnetic signals of TaAs dovetail our calculations for the 3D Weyl fermions in the strong field limit. Previous calculations only addressed $M_{||}$ over the full range of magnetic field for different types of band contacting \cite{Mikitik04prl,mikitik1996formus,Mikitik_2016}, while here we formulate a more general magnetization theory for $M_{||}$, $M_T$, and $\tau$ in an arbitrary angle for Weyl semimetals in which the band structure is also taken into account. The full version of the theory is presented in SI and here we give the conclusion in the strong field limit. The Hamiltonian for a single node of 3D massless Weyl fermions can be formulated as,
\begin{eqnarray}
\mathcal{H} &=& v_a p_x\sigma_x + v_b p_y\sigma_y + v_c p_z\sigma_z,
\end{eqnarray}
where $v_{a,b,c}$ take into account the anisotropy of the Fermi velocity, $p_{x,y,z}$ are the momentum, and $\sigma_{x,y,z}$ are the Pauli matrices. We derive the magnetization of this model in the QL (see SI) as
\begin{eqnarray}
   M_\|\propto-B\ln\Gamma \label{Eq:M_para_QL}
\end{eqnarray}
for the parallel component, and
\begin{eqnarray}
  M_{T}\propto  -B\ln\Gamma \frac{\partial \Delta}{\partial \phi}
\end{eqnarray}
for the transverse component. Here
$\Gamma$ = $2\Lambda\ell_B/  v_b\sqrt{2\Delta} $, where $-\Lambda$ is the cutoff energy of the valence band, $\Delta$ = $(v_a/v_b)\cos\alpha\cos\theta+(v_c/v_b)\sin\alpha\sin\theta$, where $\alpha=\tan^{-1}\left(\frac{v_c}{v_a}\tan\theta\right)$ and $\ell_B=\sqrt{\hbar/eB}$ is the magnetic length.

Then we check whether our calculation can well fit the experimental results for TaAs. Figure~\ref{Fig4}\textbf{A} shows that the measured $M_\parallel$ for H$\parallel$c on another sample in strong pulsed magnetic field. The data can be well fitted by using Eq. (\ref{Eq:M_para_QL}) and  Eqs. (E36) and (E37) in SI where we assume a fixed $E_F$. Our theoretical simulation of the formulas for the 3D Weyl fermions reproduces the dHvA oscillations at low fields and a significantly enhanced $M_\parallel$ in the QL. For comparison, the $M_{||}$ for the 3D massive non-relativistic electrons with parabolic energy dispersion [see Eq. (E10)] saturate in the QL.
The $M_T$ in Fig.~\ref{Fig4}\textbf{B} can be well fitted by Eq. (3) and Eqs. (E39) and (E40) in SI as well. We emphasize that the fittings based on different constraints with a fixed $N_c$ or $E_F$ give a similar trend in strong field which is distinct from the saturating magnetization for massive electrons.

The linear non-saturating magnetization in the QL is of particular interest because it is beyond Landau's theory of magnetization for classical electrons \cite{Landau30zp} (see also the classical theory of magnetization in SI).
Such feature has never been observed in topological trivial semimetals (only for bulk band) in strong magnetic field \cite{Kapitza28, Brignall74jpc, Roeland75pbc, Mcclure76jltp,Brandt77jltp}. Our calculation and experiment demonstrate the unique magnetic response in the QL for the linear energy dispersive, chiral model of the 0th LB of Weyl fermions. Recently emergent topological materials for potential applications in general manifest small Fermi surface in which the QL can be accessed in a constant magnetic field, yet their topological nature of band structure is difficult to identify by spectroscopic and electrical transport techniques. Unlike the transport properties which are difficult to model reliably, the experimental data of the magnetization is consistent with our newly developed thermodynamic model which can serve as a unique criterion for massless electrons.



\section*{Methods}
TaAs has 12 pairs of Weyl nodes which are divided into 4 pairs of W1 and 8 pairs of W2.
The $E_F$ of the as-grown single-crystalline TaAs is close to the two types of Weyl nodes which are separated 13 meV in energy space, therefore its Fermi surface consist pairs of Weyl electron pockets (see Fig. S7) \cite{zcl,arnold2016chiral}.

We prepared the single crystals of TaAs by the standard chemical vapor transfer (CVT) \cite{scheafer1964chemical,MurrayTA_mainGrowth} in this study. The large single crystals we used for the magnetic torque is shown as an inset in Fig. S1. Its polished surface shows the (001) plane which was confirmed by X-ray diffraction measurements. The single crystal for parallel magnetization measurements in the pulsed field is 1$mm$$\times$1$mm$$\times$5$mm$ for acquiring the data with higher resolution.

The magnetic torque measurements were performed using capacitive cantilever in water-cooled magnet with the steady fields up to 33T in the Chinese High Magnetic Field Laboratory (CHMFL), Hefei. In order to estimate the background signal from the cantilever and the cable, the empty cantilever was calibrated on the same conditions. The details of the calibration and measurements are shown in SI.

\begin{acknowledgments}
C.-L. Z. appreciates Lu Li's crucial comments on small angle torque theory and treats his research work as precious homework from Shui-Fen Fan. J.-L. Z. thanks Dr. Lin Jiao for sharing his analysis program with us. C.-M. W. thanks lots of discussions from Yuriy Sharlai. S.J. is supported by National Basic Research Program of China (Grant Nos. 2014CB239302) and National Natural Science Foundation of China (Grant No. 11774007). J.-L. Z. is supported by National Natural Science Foundation of China No.11504378. H.Z.L was supported by the National Key R \& D Program (Grant No.2016YFA0301700) and National Natural Science Foundation of China under Grant No. 11574127. C.-M. W. is supported by the National Natural Science Foundation of China (Grant No. 11474005). H.L. acknowledges the Singapore National Research Foundation for the support under NRF Award No. NRF-NRFF2013-03. The National Magnet Laboratory is supported by the National Science Foundation Cooperative Agreement no. DMR-1157490, the State of Florida, and the US Department of Energy. Work at Los Alamos National Laboratory was supported by the Department of Energy, Office of Science, Basic Energy Sciences program LABLF100 "Science of 100 Tesla." C.-L. Z and C.-M. W. contributed equally to this work.

\end{acknowledgments}

\bibliographystyle{unsrt}

\end{document}